\documentclass{PoS}
\usepackage{amsmath}
\usepackage{slashed}
\usepackage{multirow}

\title{The nature of the $X(3915)/X(3930)$ resonances from a coupled-channels approach}

\ShortTitle{The nature of the $X(3915)/X(3930)$ resonances}

\author{\speaker{Pablo G. Ortega}\\
        Grupo de F\'isica Nuclear and Instituto Universitario de F\'isica 
Fundamental y Matem\'aticas (IUFFyM), Universidad de Salamanca, E-37008 
Salamanca, Spain\\
        E-mail: \email{pgortega@usal.es}}

\author{Jorge Segovia\\
        Institut de F\'isica d'Altes Energies (IFAE) and Barcelona Institute of Science and Technology (BIST),
Universitat Aut\`onoma de Barcelona, E-08193 Bellaterra (Barcelona), Spain\\
        E-mail: \email{jsegovia@ifae.es}}

\abstract{The positive parity $\chi_{cJ}(2P)$ charmonium states are expected to 
lie around the 3.9 GeV/$c^2$ energy region, according to the predictions of 
quark models. 
However, a plethora of states with difficult assignment and unconventional 
properties have been discovered over the years, i.e., the $X(3872)$, $X(3940)$, 
$Y(3940)$, $X(3915)$, $X(3860)$ and the $X(3930)$ resonances, which complicates 
the description of this intriguing region.

In this work we analyze the $0^{++}$ and $2^{++}$ sectors, employing a 
coupled-channels formalism successfully applied to the $1^{++}$ sector, where 
the $X(3872)$ was described as a $D\bar D^\ast+h.c.$ molecule with a sizable 
$c\bar c$ $(2^3P_1)$ component.
This coupled-channels formalism is based on a widely-used Constituent Quark 
Model, which describes the quark-quark interactions, and the $^3P_0$ quark pair 
creation mechanism, used to couple the two and four quark sectors.

The recent controversy about the quantum numbers of the $X(3915)$ state, the 
properties of the $X(3930)$ one and the nature of the new $X(3860)$ resonance 
are analyzed in a unified theoretical framework, being all the parameters 
completely constrained from previous calculations in the low-lying heavy 
quarkonium  phenomenology.
 }

\FullConference{XVII International Conference on Hadron Spectroscopy and Structure - Hadron2017\\
		25-29 September, 2017\\
		University of Salamanca, Salamanca, Spain}

\begin{document}

\section{Introduction}

The theoretical interest in the $3.9$ GeV/c$^2$ energy region of the positive-parity charmonium spectra began in 2003, after the discovery of the $X(3872)$ by the Belle Collaboration~\cite{Choi:2003ue}. 
This state, with $J^{PC}=1^{++}$, decays through the $J/\psi\rho$ and $J/\psi\omega$ channels which are, respectively, forbidden and OZI-suppressed for a $c\bar c$ configuration.
Since then, many other structures have been observed in this energy region, like the $Y(3940)$ one, discovered by Belle~\cite{Abe:2004zs}, the $X(3940)$~\cite{Abe:2007jna} or the $X(3930)$, a $J^{PC}=2^{++}$ peak measured in 2006 by the same Collaboration in the mass spectrum of the $D\bar D$ mesons produced by $\gamma \gamma$ fusion, rapidly assigned to the $\chi_{c2}(2P)$ state at the PDG. 

Additionally, a recent controversy has appeared concerning the $X(3915)$ resonance~\cite{Uehara:2009tx,Lees:2012xs},
traditionally assigned to a $J^{PC}=0^{++}$ state. Recent works favor a $J^{PC}=2^{++}$ nature for the latter, which as a side effect would imply a large
non-$q\bar q$ component for the $X(3915)$ state~\cite{Guo:2012tv,Olsen:2014maa,Zhou:2015uva}.
Furthermore, a new charmonium-like state decaying to $D\bar D$, called $X(3860)$, has been recently reported by the Belle Collaboration~\cite{Chilikin:2017evr} 
(stimulated by the early work of Ref.~\cite{Guo:2012tv})
with a mass of $3862_{-32\,-13}^{+26\,+40}$ MeV/$c^2$ and a width of $201_{-67\,-82}^{+154\,+88}$ MeV. 
In this case, the $J^{PC}=0^{++}$ option is preferred over the $2^{++}$ one, but its quantum numbers are not determined yet. 

To study this abundance of states, all belonging to the same energy region, it is necessary to include all the nearby thresholds.
Based on previous calculations of the $J^{PC}=1^{++}$ sector by Refs.~\cite{Ortega:2009hj,Ortega:2012rs}, which used a coupled-channels 
approach including the bare $c\bar c\,(2^3P_1)$ and $D\bar D^\ast+h.c.$ components, in this work we will perform a calculation for the $0^{++}$ and $2^{++}$ channels~\cite{Ortega:2017qmg}  
including the $c\bar c\,(2^3P_{\{0,2\}})$ bare charmonium states, which are coupled to the closest thresholds. 
Recently, this formalism has also been successfully applied to the $c\bar s$~\cite{Ortega:2016mms} and $b\bar s$ sectors~\cite{Ortega:2016pgg}.

\vspace*{-.25cm}
\section{Coupled-channels model}
\label{sec:model}
\vspace*{-.25cm}

In order to describe mixed molecular-$c\bar c$ states we need two ingredients. 
On the one hand, a model for the $q-q$ interactions 
to describe the naive charmonium spectra and the hadron-hadron interactions. 
On the other hand, a formalism to describe the coupling between Fock spaces must be developed, 
through a mechanism that couples the two and four quark systems.

Regarding the first ingredient, the constituent quark model (CQM) employed in this work has been extensively 
detailed in Ref.~\cite{Vijande:2004he}, thus only its most relevant aspects will be briefly described. 
Further details of the model and explicit expressions for these interactions are given in the latter reference. 
This CQM has been successful in describing the meson~\cite{Vijande:2004he,Segovia:2008zza,Segovia:2016xqb} and
baryon phenomenology~\cite{Valcarce:2005em,Ortega:2011zza}.
The present CQM is based on the assumption that quarks acquire a 
dynamical mass as a consequence of the spontaneous chiral symmetry breaking. In order to compensate
the mass term in a chiral-invariant QCD Lagrangian we must include chiral fields. 
Then, the simplest Lagrangian that fulfills the previous properties can be written as

\begin{equation}
\label{lagrangian}
{\mathcal L}
=\overline{\psi }(i\, {\slashed{\partial}} -M(p^{2})U^{\gamma_{5}})\,\psi 
\end{equation}
where  $U^{\gamma_5}=e^{i\frac{\lambda _{a}}{f_{\pi }}\phi ^{a}\gamma _{5}}$ is 
the Goldstone boson fields matrix and $M(p^2)$ the dynamical
constituent mass. 
The Goldstone boson field matrix $U^{\gamma_{5}}$ can be expanded in terms of boson fields, 
from where one-boson exchange interaction between quarks naturally emerges.

Above the chiral symmetry breaking scale the interaction through Goldstone bosons does not act. 
However, quarks still interact through gluon exchange diagrams.
Besides, the model is completed with the confinement, a nonperturbative QCD effect.
This interaction can be modeled with a screened potential~\cite{PhysRevD.40.1653}, which takes into account the
saturation of the potential at some interquark distance due to the spontaneous creation of light-quark pairs.

About the second ingredient, the aforementioned CQM is insufficient to describe the interplay between charmonium and meson-meson channels,
hence, we need to develop a complete coupled-channels formalism. For that purpose we will follow the
formalism of Ref.~\cite{Ortega:2012rs}.
That way, the influence of the $c \bar c$ states on the dynamics
of the two meson states is studied. This allows to generate new states through 
hadron-hadron interaction, described by the coupling with heavy quarkonium states and the underlying
$q-q$ interaction.

The global hadronic state, combination of heavy quarkonium and meson-meson channels, is
\begin{equation} 
\label{ec:funonda}
 | \Psi \rangle = \sum_\alpha c_\alpha | \psi_\alpha \rangle
 + \sum_\beta \chi_\beta(P) |\phi_{M_1} \phi_{M_2} \beta \rangle
\end{equation}
being $|\psi_\alpha\rangle$ the $c\bar c$ eigenstates of the two body
Hamiltonian and 
$\phi_{M_i}$ the $c\bar n$ (or $\bar c n$)~\footnote{In this work we will denote $n$ as the light quarks $\{u,d,s\}$} eigenstates describing 
the $M_i$ mesons. Moreover, 
$|\phi_{M_1} \phi_{M_2} \beta \rangle$ is the two meson state with $\beta$ quantum
numbers coupled to total $J^{PC}$ quantum numbers
and $\chi_\beta(P)$ is the relative wave 
function between the two mesons ${M_1}{M_2}$ in the molecule. 

The description of hadron-hadron interaction from the fundamental $q-q$ interactions is obtained from
Resonating Group Method (RGM). The wave functions
of the mesons (either $c\bar c$ or $c\bar n$) 
are calculated by solving the Sch\"odinger equation for the quark-antiquark
problem using the Gaussian Expansion Method (GEM).
We employ Gaussian trial functions whose 
ranges are in geometric progression~\cite{Hiyama:2003cu}, which helps to optimize the ranges using few free parameters. 
This choice generates a dense distribution of Gaussians at small range, which is adequate for short-range potentials.

The coupling between the $c\bar c$ and the $c\bar n-n\bar c$ channels requires the creation 
of a light quark pair $q\bar q$. We will use the quark pair creation $^3P_0$ model~\cite{Micu:1968mk},
which describes the strong decay processes from simple principles, giving equivalent results compared
to microscopic calculations~\cite{PhysRevD.54.6811}. 

Accordingly, the transition potential $h_{\beta \alpha}(P)$ describing the $c\bar c\to M_1M_2$ process within the $^3 P_0$ model
can be defined as $\langle \phi_{M_1} \phi_{M_2} \beta | \mathcal{T}| \psi_\alpha \rangle = P \, h_{\beta \alpha}(P) \,\delta^{(3)}(\vec P_{\mbox{cm}})$,
being $P$ the relative momentum of the two meson state and $\mathcal{T}$ the transition operator, 
non-relativistic reduction of the $^3P_0$ pair creation Hamiltonian.

Now, the coupled-channels system for the bare charmonium states and the meson-meson channels can be expressed as an Schr\"odinger-type
equation~\cite{Ortega:2009hj} where, apart from the potential coming from $q-q$ interactions described by
the CQM, we have an effective potential coding the coupling with $c\bar c$ states,
\begin{equation}\label{ec:effPOT}
V^{\rm eff}_{\beta'\beta}(P',P;E)=\sum_{\alpha}\frac{h_{\beta'\alpha}(P')
h_{\alpha\beta}(P)}{E-M_{\alpha}}.
\end {equation}

In order to describe states above thresholds, a more suitable formalism is to describe resonances as poles in the complex $T(\vec p,\vec p',E)$ matrix, solution of 
the Lippmann-Schwinger equation. 
Employing the total potential $V_T(P',P;E)=V(P',P)+V^{\rm eff}(P',P;E)$, where $V(P',P)$ is the potential from CQM and $V^{\rm eff}_{\beta'\beta}(P',P;E)$
the effective potential of Eq.~\ref{ec:effPOT}, 
the $T$-matrix can be factorized as 
\begin{equation}
\label{ec:tmat1}
T^{\beta'\beta}(P',P;E)=T^{\beta'\beta}_V(P',P;E)
+\sum_{\alpha,\alpha'}\phi^{\beta'\alpha'}(P';E)
\Delta_{\alpha'\alpha}^{-1}(E)\phi^{\alpha\beta}(P;E) 
\end{equation}
where $T^{\beta'\beta}_V(P',P;E)$ the $T$ matrix of the CQM potential excluding the 
coupling to the $c\bar c$ pairs and $\Delta_{\alpha'\alpha}(E)=\left((E-M_{\alpha})\delta^{\alpha'\alpha}+\mathcal{G}^{\alpha'\alpha}(E)\right)$
is the propagator of the mixed state, where $M_\alpha$ is the bare mass of the 
heavy quarkonium and $\mathcal{G}^{\alpha'\alpha}(E)$ the complete mass-shift. Then, resonances emerge as zeros of the latter
propagator ($\left|\Delta_{\alpha'\alpha}(\bar{E})\right|=0$).
Additionally, the $\phi^{\beta\alpha}(P;E)$ functions are the
$^3P_0$ vertex dressed by the RGM meson-meson interaction.

\vspace*{-.25cm}
\section{Results}
\label{sec:results}
\vspace*{-.25cm}

%
%
For the present calculation, the  bare $c\bar c\,(2^3P_{\{0,2\}})$ meson states will be coupled to 
$D\bar D$ (3734 MeV/c$^2$), $\omega J/\psi$ (3880 MeV/c$^2$), $D_s\bar D_s$ (3937 MeV/c$^2$) and $D^\ast\bar D^\ast$ (4017 MeV/c$^2$) channels, 
where the threshold energies for each channel are indicated in parenthesis.
For $J^{PC}=2^{++}$ we add the $D\bar D^\ast+h.c.$ (3877 MeV/c$^2$) channel, which is closed for $0^{++}$ sector. The $D^\ast \bar D^\ast$ channel is included because it is the only threshold 
which couples to the $c\bar c\,(2^3P_0)$ in $S$ wave and, hence, its effect is expected to be essential to describe the $0^{++}$ sector. Furthermore, Heavy Quark Spin Symmetry tells us that
the $D^\ast\bar D^\ast$ in the $2^{++}$ sector has the same dynamics as the $D\bar D^\ast+h.c.$ channel in the $1^{++}$ sector, which is fundamental in describing the $X(3872)$.

Table~\ref{tab:model A} shows the masses and widths of the states found in the coupled-channels calculation 
using the original parameters of Ref.~\cite{Ortega:2009hj} (denoted as {\it model A}).
In addition, to explore the sensitivity of the results, a second calculation (named {\it model B}) is shown in the same table, where the bare mass of 
the $c\bar c\,(2^3P_{\{0,2\}})$ pairs has been slightly changed ($0.25\%$) and the value of the $^3P_0$ strength parameter $\gamma$ has been taken from Ref.~\cite{Segovia:2012cd}. 
Such changes in the initial parameters are within the uncertainties of the model.

\begin{table*}[!t]
\begin{tabular}{cccccccccc}
Model &$J^{PC}$ & Mass & Width & ${\cal P} [c\bar{c}] $ & ${\cal P}[D\bar D]$ & ${\cal P}[D \bar D^{\ast}]$ & ${\cal 
P}[\omega J/\psi]$ & ${\cal P}[D_s\bar D_s]$ & ${\cal P}[D^{\ast} \bar D^{\ast}]$\\
\hline
\multirow{3}{*}{A}&$0^{++}$ & $3890.3$ & $6.7$ & $44.1\%$ & $21.6\%$ & $-$ & $28.4\%$ & $2.6\%$ & $3.3\%$ \\
&$0^{++}$ & $3927.4$ & $229.8$ & $19.2\%$ & $66.3\%$ & $-$ & $5.3\%$ & $3.7\%$ & $5.5\%$ \\
&$2^{++}$ & $3925.6$ & $19.0$ & $42.2\%$ & $11.3\%$ & $37.0\%$ & $4.0\%$ & $0.4\%$ & $5.1\%$ \\
\hline
\multirow{3}{*}{B}&$0^{++}$ & $3889.0$ & $11.8$ & $43.5\%$ & $27.3\%$ & $-$ & $20.4\%$ & $3.8\%$ & $4.9\%$ \\
&$0^{++}$ & $3947.5$ & $201.6$ & $19.4\%$ & $66.0\%$ & $-$ & $3.7\%$ & $8.0\%$ & $2.9\%$ \\
&$2^{++}$ & $3915.1$ & $19.8$ & $37.8\%$ & $14.1\%$ & $36.4\%$ & $5.12\%$ & $0.4\%$ & $6.1\%$ \\
\end{tabular}
\caption{\label{tab:model A} Mass and decay width, in MeV, and probabilities of 
the different channels, for models A and B.}
\end{table*}

The same number of states is predicted by both models: two resonances with $J^{PC}\!\!=\!0^{++}$, a broad and a narrow one, and  
one with $J^{PC}\!\!=\!2^{++}$. The mass and width of the $J^{PC}\!\!=\!2^{++}$ state is compatible with those measured for the $X(3930)$ for model A, but its mass
moves closer to the mass region of the $X(3915)$ for model B, so its assignment remains unclear. 

In the $J^{PC}\!\!=\!0^{++}$ region both models show a similar phenomenology. On the one hand, the mass of the narrow state is more compatible with the new $X(3860)$ resonance than to the $X(3915)$. 
However, its predicted width is smaller than the experimental one, which is most likely connected to the position of the node in the $c\bar c\,(2^3P_0)$ bare wave function, which makes the width
sensitive to small variations in the wave function structure or, equivalently, to the mass of the $X(3860)$. 

The second $J^{PC}\!\!=\!0^{++}$ state is more intriguing. Its mass suggests a preliminary assignment to the $Y(3940)$ resonance, but, as already mentioned, its width is far from the experimental value. 
Alternatively, one can claim the second $0^{++}$ state is actually the $X(3860)$, as it is a broad state whose width agrees with the experimental one and its mass is close to the experiments, whose value reaches up to $3900$ MeV. Within this picture, the first state seems too narrow, which can complicate its measurement in the experiments of Ref.~\cite{Chilikin:2017evr}.
What appears to be clear is that none of the $J^{PC}\!\!=\!0^{++}$ states seems to favor a $X(3915)$ assignment, neither by mass nor by width.

Nevertheless, the predicted $J^P=2^+$ state is compatible with both $X(3915)$ and $X(3930)$, within the uncertainties of the CQM and the value of the $^3P_0$ quark-pair creation parameter. 
So, in order to further analyze the properties of the $X(3915)$ resonance, we can explore its decay properties. Under a $J^{PC}\!\!=\!0^{++}$ hypothesis for the $X(3915)$ state we are unable to describe the experimental branching fractions. In contrast, the product of the two-photon decay width and the branching fraction to $\omega J/\psi$ and 
$D\bar D$ channels can be calculated assuming the $X(3915)$ and $X(3930)$ are two faces of the same $J^{PC}\!\!=\!2^{++}$ resonance. 
Such results are quoted in Table~\ref{tab:BR J2}, where the decay to the $D\bar D^*$ channel is also shown.
The results for both models A and B are very similar, not far from the experimental values. 
Consequently, our results suggest that both the $X(3915)$ and $X(3930)$ resonances fit within a $J^{PC}\!\!=\!2^{++}$ assignment, pointing to the conclusion that they 
are two faces of the same state, which agrees with Ref.~\cite{Branz:2010rj}.

\begin{table*}[!t]
\begin{tabular}{ccccc}
 & Belle & BaBar & model A &  model B  \\
\hline
$\Gamma_{\gamma \gamma} \times {\cal B}(2^{++}\to \omega J/\psi )$ & $18(5)(2)$ & $10.5(1.9)(0.6)$ & $20.9$ & $24.9$  \\
$\Gamma_{\gamma \gamma} \times {\cal B}(2^{++}\to D\bar D )$ & $180(50)(30) $ & $249(50)(40)$ & $75.4$ & $81.4$  \\
$\Gamma_{\gamma \gamma} \times {\cal B}(2^{++}\to D\bar{D^\ast})$ & - & - & $196.0$ & $151.9$  \\
\end{tabular}
\caption{\label{tab:BR J2} Product of the two-photon decay width and the branching fraction (in eV) for the 
$J^{PC}=2^{++}$ sector for each model, compared with Belle and BaBar Collaboration experimental results~\cite{Uehara:2009tx,Lees:2012xs,Uehara:2005qd,Aubert:2010ab}}
\end{table*}

\vspace*{-.35cm}
\section*{Acknowledgments}
\vspace*{-.35cm}

This work has been partially funded by EU's Horizon 2020 programme under the Marie Sk\l{}odowska-Curie grant agreement no. 665919; 
by Spanish MINECO under Contract nos. FPA2014-55613-P and SEV-2016-0588;  
and by Junta de Castilla y Le\'on and ERDF under Contract no. SA041U16.

\vspace*{-.35cm}
\bibliographystyle{JHEP}
\bibliography{X3915}

\end{document}